\documentclass[10pt, reprint, twocolumn, nofootinbib, showkeys, aps, prd, colorlinks=true, linkcolor=blue, urlcolor=blue, citecolor=red, anchorcolor=blue]{revtex4-2}
\usepackage[T1]{fontenc}
\usepackage{amsmath}
\usepackage{amsfonts,txfonts,pxfonts}
\usepackage{bm}% For bold math (recommended for Greek)
\usepackage{amssymb}
\usepackage{orcidlink}
\usepackage[export]{adjustbox}
\usepackage{fix-cm}
\usepackage{adjustbox}
\usepackage{latexsym,amsmath,amssymb}
\RequirePackage{graphicx}
\RequirePackage{mathptmx}
\usepackage[arrowdel]{physics}
\usepackage{graphicx,epstopdf}
\usepackage[scr=rsfs]{mathalpha}
\usepackage{slashed}
\usepackage{bm}
\usepackage{cases}
\usepackage{color}
\usepackage{bm}
\usepackage{dcolumn}
\usepackage{lipsum}
\usepackage{hyperref}
\begin{document}
\title{Gravity theory of the \textit{generalized mass-to-horizon entropy}: The Iyer–Wald approach}
\author{Subhra Mondal \orcidlink{0009-0003-6469-6238}}
 \email{cosmology313@gmail.com}
\author{Amitava Choudhuri \orcidlink{0000-0001-9499-8585}}
 \email{amitava\_ch26@yahoo.com}
\affiliation{Department of Physics, \href{https://ror.org/05cyd8v32}{The University of Burdwan}, Golapbag, Purba Bardhaman - 713104, West Bengal, India.}
%%%%%%%%%%%%%%%%%%%%%%%%%%%%%%%%%%%%%%%%%%%%%%%%%%%%%%%%%%%%%%%%%%%%%%%%%%%%%%%
\hspace{1cm}
\begin{abstract}
In this letter, our aim is to study the gravitational origin of the \textit{generalized mass-to-horizon entropy} (GMHE), described by an entropic exponent index $n$ and a multiplicative parameter $\gamma$. We employ Iyer-Wald's approach within the modified $\mathfrak{f(R)}$ gravity scenario in order to assess the horizon entropy of the black hole (BH) by considering solutions with constant curvature for the spherically symmetric vacuum field equations. In this process, we explicitly show that the GMHE can effectively be reconstructed from a generalized Lagrangian $\mathfrak{L}\propto \mathfrak{R}^{1+\epsilon}$, where $\epsilon\equiv\frac{n-1}{2}$ measures tiny departures from the standard formulation of general relativity (GR). Finally, we discuss the physical implications of our results in relation to cosmology and the prospects for alleviating the thermodynamic instability of Schwarzschild BHs. 
\end{abstract}

\maketitle
%\tableofcontents
\section{Introduction}\label{Intro}
\par The Boltzmann-Gibbs (BG) statistical mechanics stands as one of the most successful and robust frameworks in modern physics. Its applications are incredibly broad, ranging from standard thermodynamics and solid-state physics to high-energy physics and information theory, in both classical and quantum sectors~\cite{landau2013statistical}. Despite its huge success, a closer look reveals severe difficulties or anomalies in describing the non-standard or complex systems, highlighting its limited applicability (see Refs.~\cite{tsallis1998role,ogorodnikov1957statistical,lynden1967statistical}). Gibbs~\cite{gibbs1902elementary} himself showed that systems with divergent canonical partition functions, such as gravitational systems (long-range interactions), violate the BG theory. It is important to note that the standard BG statistics, which leads to the famous Bekenstein-Hawking area law~\cite{bekenstein1973black,hawking1974black} for BHs, break down in explaining thermodynamic extensivity if BHs are treated as ($1+3$)-dimensional objects~\cite{hooft1985quantum,susskind1993string,strominger1996microscopic}. 
These clear conceptual boundaries have driven physicists to look beyond standard models, searching for broader frameworks that can handle systems where standard thermodynamics and statistical mechanics fail.
\par The main problematic issue with the concept of BH entropy is due to the unavailability of a proper description in conventional statistical mechanics. Rather, it depends on Bekenstein and Hawking's definition, which is nonadditive and nonextensive in nature. As a result, gravity-related applications cannot be well defined by the laws of standard thermodynamics and statistical mechanics. In order to resolve this difficulty, some nonextensive statistical mechanical formulations have been developed and employed to explain phenomena in BHs and other cosmological applications (see~\cite{ccimdiker2023equilibrium,nojiri2021area,nojiri2022nonextensive,nojiri2022alternative,nojiri2022barrow,promsiri2020thermodynamics,promsiri2021solid,tannukij2020thermodynamics,nakarachinda2021effective,czinner2025hawking,nakarachinda2025thermodynamics,saridakis2020barrow,dkabrowski2020geometrical,mondal2024dynamics,mondal2025dynamics,komatsu2017cosmological,komatsu2016general,nunes2016probing,liu2022spectrum,majhi2017non,luciano2021q,di2022sign,di2022barrow,asghari2022observational,abreu2022statistical}, and the references therein). These nonextensive statistics/entropies have become popular nowadays and include a variety of propositions, such as Tsallis statistics~\cite{tsallis1988possible,tsallis2009introduction}, Sharma–Mittal statistics~\cite{sharma1975new,sharma1977new,masi2005step}, R\'{e}nyi statistics~\cite{renyi1961measures} and Kaniadakis statistics~\cite{kaniadakis2002statistical,kaniadakis2005statistical,kaniadakis2001non}, the Tsallis–Cirto entropy~\cite{tsallis2013black,tsallis2019blackholeentropy}, Zamora-Tsallis entropy~\cite{zamora2022thermodynamically1,zamora2022thermodynamically2} and Barrow entropy~\cite{barrow2020area} are some specific examples. The application of nonextensive entropies to self-gravitating systems is evident due to the significant impact that long-range interactions play in nonextensive thermodynamics. Nonextensivity that results from long-range interactions is anticipated in Tsallis's statistical mechanics. R\'{e}nyi statistics has been extensively used in quantum information theory, and R\'{e}nyi entropy quantifies the entanglement of quantum mechanical systems. Sharma–Mittal statistics simply extend the ideas of Tsallis and R\'{e}nyi statistics, while the Lorentz transformation of special relativity inspires building Kaniadakis statistics. It is worth mentioning that some of these entropies have formal similarities. However, they differ in their conceptual foundations. For instance, Tsallis-Cirto entropy arises from the requirement of the extensive property of classical BH thermodynamics, whereas Barrow entropy emerges from consideration of the fractal nature of the BH horizon, motivated by the structure of the COVID-$19$ virus, where the effects of quantum gravitational spacetime foam have been taken care of. Zamora-Tsallis entropy arises in the cosmological context by considering a thermodynamically consistent entropic-force model, where the Legendre structure of thermodynamics has been conserved.
\par Another promising formalism in pursuit of nonextensive statistics is the entropic formalism based on \textit{generalized mass-to-horizon ratio}~\cite{gohar2024generalized,denkiewicz2026mass,gohar2024foundations,sheykhi2025emergence}. Recently, the GMHE framework has been explored in different contexts, especially in the cosmological sector~\cite{gohar2024generalized,denkiewicz2026mass,gohar2024foundations,luciano2025modified,luciano2025modified_Cos,basilakos2025modified,luciano2026baryogenesis,sheykhi2025emergence,ali2025growth,mondal2026cosmological}. An analysis presented in Ref. \cite{gohar2024generalized} reveals that this framework can create a cosmological model that is in agreement with observation, quite similarly to the widely accepted Lambda cold dark matter ($\Lambda$CDM) model, provided specific conditions. In Ref~\cite{gohar2024foundations}, the authors study the foundations of entropic cosmologies, addressing their inconsistencies, potential solutions, and identifying signs of dead ends. In Ref.~\cite{denkiewicz2026mass}, late-time acceleration of our Universe has been investigated along with observational data sets and claim that the GMHE-inspired cosmological model is statistically preferred over the standard $\Lambda$CDM profile. The study in Ref. \cite{sheykhi2025emergence} formulates the modified Friedmann equations based on GMHE and searches for the origin of cosmic space by exploiting the equilibrium approach, inspired by Padmanabhan's idea of `emergent gravity' and the first law of thermodynamics on the apparent horizon. Authors in Ref. \cite{basilakos2025modified} derive the modified Friedmann equations from this entropy and add an effective DE component by using the gravity-thermodynamics conjecture. Their findings, based on different datasets, show that this modified model is consistent with observations. The study of the generation of the baryon asymmetry within this cosmological setting has been explored in Ref. \cite{luciano2026baryogenesis} and establishes certain limits on the entropic parameters. In Ref. \cite{luciano2025modified}, authors impose observational constraints on this model, and note this model closely resembles the $\Lambda$CDM profile without significant deviations. The effect of this modified cosmology on the growth of matter density perturbation and primordial gravitational waves in the early Universe has been analyzed in Ref. \cite{luciano2025modified_Cos}. In Ref.~\cite{ali2025growth}, background cosmology and growth of density perturbation in the context of GMHE have been investigated, which is found to be fully compatible with observed data. In our previous article~\cite{mondal2026cosmological}, we explored the cosmological dynamics and structure formation by employing the spherical top-hat collapse formalism in the same cosmological setup. 
\par Although a lot of studies of GMHE in the field of cosmology exist in the literature, its application to BH is still lacking. In this work, we aim to investigate the roots of the GMHE of the BH horizon in gravity theory. It is established that the entropy associated with a BH horizon in any diffeomorphism-invariant gravity theory can, theoretically, be obtained from the effective Lagrangian of the system through the Noether charge method proposed by Iyer and Wald~\cite{wald1993black,iyer1994some,iyer1995comparison}. In analogy with the recently established study on the Lagrangian approach to Tsallis entropy via the Iyer-Wald approach~\cite{d2024lagrangian}, we seek an extension of GR to GMHE for BHs at the fundamental level. The consistency of the Iyer-Wald formula can, in principle, be verified by reproducing the standard area law in GR, whereas corrections do emerge for any modified gravity scenario that drifts from the Einstein-Hilbert action. Therefore, whether any entropic theory developed from first principles that differs from the standard Bekenstein-Hawking case, one can exploit the Iyer-Wald mechanism to derive a modified Lagrangian for the corresponding entropic formula~\cite{Hammad_modified}. Utilizing this idea, we demonstrate that the GMHE emerges from a specific category of $\mathfrak{f(R)}$ theory of gravity.
\par The remainder of this letter is organized as follows. In Secs.~\ref{GMHE} and~\ref{IW_formalism}, we outline the theoretical background of the GMHE and the Iyer-Wald formalism in the context of BH, respectively. In Sec.~\ref{sec:f(R)_BHs}, we provide a brief overview of the spherically symmetric solutions of BH in the $\mathfrak{f(R)}$ modified gravity scenario. Utilizing the Iyer-Wald approach and considering small deviations from GR, we derive the effective Lagrangian that can reconstruct GMHE. Cosmological implications and BH thermodynamic stability have been discussed in Sec.~\ref{PhysImpl}. Sec.~\ref{Concl} is dedicated to the summary and conclusions.
%%%%%%%%%%%%%%%%%%%%%%%%%%%%%%%%%%%%%%%%%%%%%%%%%%%%%%%%%%%%%%%%%%%%%%%%%%%%%%%%%%%%%%%%%%%%%%%%%%%%%%%%%%%%%%%%%%%%%%%%%%%%%%%%%%%%%%%
\section{Generalized Mass-to-Horizon Entropy}
\label{GMHE}
\par The entropy of a BH quantifies the information regarding the inside of a BH, which an outside observer cannot access. The standard Bekenstein-Hawking entropy $S_{BH}$ measures entropy proportional to the surface area $\mathcal{A}_{hor}$ of the BH horizon\footnote{Throughout this article, we assume a unit system in which $k_B=c=\hbar=1$, unless stated otherwise. Common physics notations have been used for fundamental physical constants. Additionally, we adopt a spacetime signature $(-,+,+,+)$ for ($1+3$)-dimentional systems.}
\begin{equation}
    S_{BH}=\frac{\mathcal{A}_{hor}}{4G}\,,
\end{equation}
rather than its volume, in accordance with the holographic principle~\cite{hooft2001holographic,susskind1995world,bousso2000holographic,bousso2002holographic}. It suggests that the degrees of freedom within the bulk can be correlated with the two-dimensional boundary of the BH. The number of fundamental degrees of freedom in a closed spatial region is finite and cannot exceed the Bekenstein-Hawking limit. At the same time, the Hawking temperature defines the temperature in terms of surface gravity associated with the horizon. For thermodynamically consistent entropic formulations, including the Bekenstein-Hawking one, are based on the following three pillars~\cite{gohar2024generalized,gohar2024foundations,sheykhi2025emergence}: 
\begin{itemize}
\item \textbf{Holographic connection:} The BH horizon must possess properties like energy $E$, mass $M$, entropy $S$, and temperature $T$ (Hawking temperature).
\item \textbf{Clausius relation:} These properties are associated through the Clausius relationship: $dE = dM = T dS$.
\item \textbf{Geometric correspondence
:} A linear relationship between mass and the horizon radius $r_{hor}$, known as the standard mass to horizon ratio (SMHR), must be obeyed: $M = \frac{1}{G} r_{hor}$.
\end{itemize}
However, the Clausius relation leads to an inconsistent mass-energy relationship when the Hawking temperature is applied together with nonextensive entropies on the BH horizon. This indicates that, apart from Bekenstein-Hawking entropy, the Hawking temperature is inconsistent with other nonextensive entropies as they feature fundamental thermodynamic incompatibility~\cite{nojiri2021area,ccimdiker2023equilibrium}. In order to apply nonextensive entropies along with the holographic principle on the horizons instead of the Bekenstein-Hawking entropy, Gohar and Salzano ~\cite{gohar2024generalized} proposed a nonlinear generalized mass to horizon ratio (GMHR)~\cite{gohar2024generalized,luciano2025modified,basilakos2025modified,luciano2025modified_Cos,sheykhi2025emergence}
\begin{equation}\label{GMHR}
    M = \frac{\gamma}{G} r^n_{hor}\, ,
\end{equation}
which, by applying alongside the Hawking temperature, one can obtain a novel, thermodynamically consistent two-parameter extension of Bekenstein-Hawking horizon entropy~\cite{gohar2024generalized,luciano2025modified,luciano2025modified_Cos,basilakos2025modified,luciano2026baryogenesis,sheykhi2025emergence}
\begin{equation}\label{GS_entropy_area}
S_{GS} = \gamma\,\frac{2\pi}{G}\,
\frac{n}{n+1}\left(\frac{\mathcal{A}_{hor}}{4\pi}\right)^{\frac{n+1}{2}}= \gamma\,\frac{2n}{n+1} r\,^{n-1}_{hor}S_{BH}\, ,
\end{equation}
aligning with the principle of holography. Over here, $n$ denotes a positive real dimensionless entropic exponent parameter, while the non-negative multiplicative parameter \footnote{We consider $\gamma(n) = L_\gamma^{1-n}$, where $L_\gamma$ is a typical length scale.}$\gamma$ has dimensions of $[L]^{1-n}$. In particular, for $\gamma =n =1$, the SMHR, along with the standard Bekenstein-Hawking entropy $S_{BH}$, can be retrieved. It is worth highlighting that we now possess sufficient versatility to derive various forms of entropy beyond the standard case. For example, by taking $n = 2\delta - 1$, we get the nonextensive Tsallis-Cirto entropy \cite{tsallis2013black,tsallis2019blackholeentropy,mondal2024dynamics} and Zamora-Tsallis entropy \cite{zamora2022thermodynamically1,zamora2022thermodynamically2} when $n = d - 1$. Similarly, quantum-corrected Barrow entropy \cite{barrow2020area,mondal2025dynamics} can be obtained for $n = 1 + \Delta $\,. Here $\delta$, $d$, and $\Delta$ are parameters for the respective entropic theories.
%%%%%%%%%%%%%%%%%%%%%%%%%%%%%%%%%%%%%%%%%%%%%%%%%%%%%%%%%%%%%%%%%%%%%%%%%%%%%%%%%%%%%%%%%%%%%%%%%%%%%%%%%%%%%%%%%%%%%%%%
\section{The Iyer-Wald Formalism}\label{IW_formalism}
\par It is well known that for any diffeomorphism-invariant Lagrangian of a general classical gravity theory with arbitrary matter fields in $D$ spacetime dimensions, the conventional BH thermodynamics can be obtained by invoking the Iyer-Wald formalism~\cite{wald1993black,iyer1994some,iyer1995comparison}. It is a covariant phase-space formulation that constructs conserved Noether charges and entropy functionals via the Lagrangian approach in diffeomorphism-invariant gravity theories and admits BH solutions with a bifurcate Killing horizon. In this approach, the standard Bekenstein-Hawking entropy can be derived by employing the Einstein-Hilbert Lagrangian of GR. However, in modified gravity theories, this induces corrections to the background spacetime curvature~\cite{briscese2008black,vollick2007noether,momeni2025wald,d2024lagrangian}. This sounds interesting and might help us to find the underlying gravity rules of any entropic gravity models, including the GMHE-inspired modified gravity scenario that deviates from the standard Bekenstein-Hawking case. 
\par In this article, we intend to implement the Iyer-Wald formalism to GMHE-inspired modified gravity. To do this, we first outline the formalism for our current purpose. According to Noether's theorem, for any gravity theory with dynamic field variable $\phi \equiv \{\mathfrak{g_{ab}},\psi\}$ characterized by a diffeomorphism-invariant Lagrangian $\mathfrak{L}(\phi)$, one can find a Noether $(D-1)$-form current $\bm{J}$ and $(D-2)$-form charge $\bm{Q}$. The variation of Lagrangian $D$-form $\bm{L}=\mathfrak{L}(\phi) dx^0 \wedge \dots \wedge dx^{D-1}$ associated with the variation in field variable $\delta\phi$ takes the form~\cite{wald1993black,iyer1994some,iyer1995comparison}
\begin{equation}
    \delta\bm{L} = \bm{E} \delta\phi + d\bm{\Theta}(\phi, \delta\phi)\,,
\end{equation}
where $\bm{E}$ form denotes equations of motion of the fields and and $\bm{\Theta}$ is the $(D-1)$-form symplectic potential. The familiar Noether $(D-1)$-form current associated with the diffeomorphisms generated by the arbitrary vector field $\xi = \xi^a \partial_a$
\begin{equation}
    \bm{J} = \bm{\Theta}(\phi, \mathcal{L}_\xi \phi) - \xi \cdot \bm{L}\,,
\end{equation}
where $\mathcal{L}_\xi \phi$ represents the Lie derivative of $\phi$ along the Killing vector field $\xi$. When the classical equations of motion are satisfied, i.e. $\bm{E}=0$, the conserved Noether current $(D-1)$-form simplifies to $\bm{J} = d\bm{Q}$. The current flux $\bm{J}$ passing through the bifurcation surface of a BH horizon $\Sigma$ reflects the entropy exchange between two spacetime regions that $\Sigma$ separates. For perturbations to nearby stationary states, one can naturally find the traditional form of the first law of thermodynamics of BH 
\begin{equation}
 \delta\int_{\Sigma} \bm{Q} = \delta\mathcal{E} - \Omega_{hor}^{(\mu)}\,\delta\mathfrak{J}_{(\mu)} = \frac{\kappa}{2\pi}\delta S \,,
\end{equation}
by choosing $\xi$ to be the symmetry of the dynamical fields. This uniquely identifies the BH entropy as a local geometrical quantity proportional to the integral of the Noether charge at the bifurcation surface of the BH horizon~\cite{wald1993black,iyer1994some,iyer1995comparison,jacobson1994black}
\begin{equation}
    S = \frac{2\pi}{\kappa} \int_{\Sigma} \bm{Q}\,.
\end{equation}
Here $\mathcal{E}$ and $\mathfrak{J}_{(\mu)}$ denote the canonical energy and angular momentum, respectively. $\Omega_{hor}^{(\mu)}$ represents the angular velocity of the horizon, and $\kappa$ refers to the BH surface gravity.
For a general gravitational theory described by a diffeomorphism-covariant Lagrangian $D$-form~\cite{iyer1994some}
\begin{widetext}
\begin{equation}
   \bm{L} \equiv \bm{L}(\mathfrak{g_{ab}, R_{abcd}, \nabla_{a_\text{1}} R_{bcde}, \dots,\nabla_{(a_\text{1}}\dots\nabla_{a_m)} R_{bcde},\psi,\nabla_{a_\text{1}}\psi,\nabla_{(a_\text{1}}\dots\nabla_{a_{\ell})}\psi})\,,
\end{equation}
\end{widetext}
the Iyer-Wald formulation of entropy yields
\begin{equation}\label{Iyer-Wald entropy}
    S = -2\pi \int_{\Sigma}\, \frac{\delta\mathfrak{L}}{\delta\, \mathfrak{R_{abcd}}}\mathfrak{\epsilon_{ab} \epsilon_{cd}} \sqrt{\mathfrak{h}}\, d^{D-2}x\,,
\end{equation}
where $\mathfrak{R_{abcd}}$ refers to the Riemann curvature tensor, and $\mathfrak{\epsilon_{ab}}$ serves as the binormal to $\Sigma$, satisfying the conditions $\mathfrak{\epsilon_{ab} = -\epsilon_{ba}}$ and $\mathfrak{\epsilon_{ab} \epsilon^{ab}} = -2$. $\sqrt{\mathfrak{h}}\, d^{D-2}x$ is the elementary surface area on the horizon crossection.
%%%%%%%%%%%%%%%%%%%%%%%%%%%%%%%%%%%%%%%%%%%%%%%%%%%%%%%%%%%%%%%%%%%%%%%%%%%%%%%%%%%%%%%%%%%%%%%%%%%%%%%%%%%%%%%%%%%%%%%%%
\section{Generalized Mass-to-Horizon Lagrangian from \texorpdfstring{$\mathfrak{f(R)}$}{f(R)} gravity}\label{sec:f(R)_BHs}
\par Let us consider a diffeomorphism-invariant theory of gravity in $4$ spacetime dimensions. We shall assume that the gravity Lagrangian is $\mathfrak{f(R)}$, which is a modified form of the Einstein-Hilbert Lagrangian. By adding matter term in $\mathfrak{f(R)}$ theory of gravity, the total action is given by~\cite{sotiriou2010f,de2010f}
\begin{equation}\label{fR_reconstructed action}
    I=\dfrac{1}{16\pi G}\int \mathfrak{f(R)}\, \sqrt{-\mathfrak{g}}\,d^4x + \int \mathcal{L}_M(\phi) \, d^4 x \, ,
\end{equation}
where $\mathfrak{f(R)}$ is a functional form of the Ricci scalar curvature, $\mathfrak{R=g_{ab}R^{ab}}$. $\mathfrak{R_{ab}}$ is the covariant Ricci tensor components, and $\mathfrak{g}$ is the determinant of the covariant space-time metric tensor components $\mathfrak{g_{ab}}$. $\mathcal{L}_M(\phi)$ is the matter Lagrangian that depends on dynamic field variable $\phi$.
\par In metric formalism, the field equation can be derived by varying the action with respect to the metric, which yields
\begin{equation}\label{eq:FE}
    \mathfrak{f'(R)R_{ab}}-\frac{1}{2}\mathfrak{g_{ab}f(R)-(\nabla_a\nabla_b-g_{ab}\Box)f'(R)}= 8\pi G\, \mathfrak{T_{ab}}\,, 
\end{equation}
where 
\begin{equation}
    \mathfrak{T_{ab}}\overset{\text{def}}{=} - \frac{2}{\sqrt{-\mathfrak{g}}}\frac{\delta \mathcal{L}_M}{\delta \mathfrak{g^{ab}}}\, ,
\end{equation}
stands for covariant components of the energy-momentum tensor of matter fields. $\mathfrak{\nabla_a}$ is the covariant derivative operator, and $\Box\equiv \mathfrak{g_{ab}\nabla^a\nabla^b}$ represents the D'Alembert operator. One can note, for $\mathfrak{f(R)=R}$, Eq.~\eqref{eq:FE} reduces to the field equation in GR, as expected. Contracting
the indices in Eq.~\eqref{eq:FE} with metric, provides us
\begin{equation}
    \mathfrak{f'(R)R}-2\mathfrak{f(R)}+3\Box \mathfrak{f'(R)}= 8\pi G \, \mathfrak{T}\,,
    \label{eq:trace}
\end{equation}
where $\mathfrak{T = g^{ab}T_{ab}}$ is the trace of the energy-momentum tensor.
\par We are interested in time-independent, spherically symmetric solutions of the vacuum spacetimes ($\mathfrak{T_{ab}} = 0$ and $\mathfrak{T} = 0$) field equations. From properties of maximally symmetric subspaces, the spacetime line element reads 
\begin{equation}
    ds^2=-h(r)dt^2+\dfrac{1}{p(r)}dr^2+r^2\left(d\theta^2+\sin^2\theta\, d\phi^2\right)\,.
    \label{eq:metric}
\end{equation}
The metric functions $h(r)$ and $p(r)$ are not unique, but rather depend on the particular type of BH solution. In particular, one might search for constant curvature solutions $\mathfrak{R=R}_0\in\mathbb{R}$ at a de-Sitter point corresponding to a vacuum solution, obeying\footnote{In this context, it is worth noting that Eq.~\eqref{eq:const_sol} permits two types of solutions: $\mathfrak{f(R)}\propto\mathfrak{R}=$ constant, and $\mathfrak{f(R)}\propto \mathfrak{R}^2$~\cite{briscese2008black,de2009black,de2009blackErratum}. For simplicity and convenience, we assume a solution with constant curvature $\mathfrak{R}=\mathfrak{R}_0$ uniformly across all spacetimes.} 
\begin{equation}
    \mathfrak{f'(R}_0)\mathfrak{R}_0 -2\mathfrak{f(R}_0)=0\,.
    \label{eq:const_sol}
\end{equation}
In this case, one finds~\cite{multamaki2006spherically}
\begin{equation}
\label{hsol}
    h(r)=p(r)=1-\frac{2GM}{r}-\dfrac{\mathfrak{R}_0 r^2}{12}\,.
\end{equation}
This characterizes the Schwarzschild-de Sitter solution of BH derived from the Lagrangian density $\mathfrak{f(R)}=\mathfrak{R}-2\Lambda$, where $\Lambda>0$ denotes the cosmological constant. Here, $\mathfrak{R}_0$ mimics the cosmological constant through the identification $\mathfrak{R}_0=4\Lambda$.
\par The horizon radius, $r_{hor}$ satisfies the condition $h(r_{hor})=p(r_{hor})=0$. In this scenario, considering small but non-zero $\mathfrak{R}_0$, we can perform a first-order perturbative expansion in powers of $\mathfrak{R}_0$ around the unperturbed Schwarzschild radius $r_S\equiv 2G M$ to derive
\begin{equation}
\label{rh}
    r_{hor}= r_S\left(1+\frac{\mathfrak{R}_0 }{12}r_S^2\right)+\mathcal{O}(\mathfrak{R}_0^2)\,,
\end{equation}
where $\mathfrak{R}_0\rightarrow0$ retrieves $r_{hor}=r_S$. Consequently, the area of the event horizon yields
\begin{equation}
    \mathcal{A}_{{hor}}=4\pi r_{hor}^2=4 \pi r_S^2 \left(1+\frac{\mathfrak{R}_0}{6} r_S^2\right)+\,\mathcal{O}(\mathfrak{R}_0^2)\,.
    \label{eq:area}
\end{equation}
\par The Iyer-Wald entropy formula~\eqref{Iyer-Wald entropy} can be applied to specified choices of diffeomorphism-invariant gravity theories. For a $\mathfrak{f(R)}$ theory of gravity in $4$ spacetime dimentions with empty spacetimes, Eq.~\eqref{Iyer-Wald entropy} can be recast as~\cite{vollick2007noether,briscese2008black}
\begin{equation}\label{entropy_f(R) horizon}
    S=\dfrac{\mathcal{A}_{hor}}{4G}\left[\mathfrak{f'(R)}\right]_{\Sigma}\,.
\end{equation}
Let us now apply Eq.~\eqref{entropy_f(R) horizon} to the BH horizon in a $\mathfrak{f(R)}$ gravity theory. For that purpose, we aim to evaluate $\mathfrak{f'(R)}=\frac{\partial\mathfrak{f(R)}}{\partial \mathfrak{R}}$ on the event horizon of BH for the entropy~\eqref{GS_entropy_area} under consideration. As we have mentioned $\mathfrak{R}=\mathfrak{R}_0$ everywhere across all spacetimes for our case, the formula~\eqref{entropy_f(R) horizon} reads 
\begin{equation}\label{entropy_f(R) R=R_0}
    S=\dfrac{\mathcal{A}_{hor}}{4G}\left[\mathfrak{f'(R)}\right]_{\mathfrak{R}=\mathfrak{R}_0}\,.
\end{equation}
Comparing Eq.~\eqref{entropy_f(R) R=R_0} with Eq.~\eqref{GS_entropy_area} and employing Eq.~\eqref{eq:area}, we get
\begin{equation}
\label{fprime_expanded}
\mathfrak{f'(R)}
= \frac{2\gamma\,n}{n+1}\left[4G^2M^2\left(1+\frac{2}{3}G^2M^2\mathfrak{R}\right)\right]^\frac{n-1}{2}\,.
\end{equation}
After integrating over $\mathfrak{R}$, one obtains
\begin{equation}\label{fR_reconstructed}
\mathfrak{f(R)} = c_1+\frac{24\gamma\,n}{(n+1)^2}(2GM)^{n-3}\left(1+\frac{2}{3}G^2M^2\mathfrak{R}\right)^\frac{n+1}{2}\,.
\end{equation}
The arbitrary integration constant $c_1$ can be determined by assessing Eq.~\eqref{fR_reconstructed} under the standard Bekenstein-Hawking entropic condition: 
\begin{equation}
   \mathfrak{f(R)}\Big|_{\gamma=n=1}=c_1+\mathfrak{R}+\frac{3}{2 G^2 M^2}\,.
\end{equation}
This expression reproduces the Schwarzschild-de Sitter solution, conforming to $\mathfrak{f(R)}=\mathfrak{R}-2\Lambda$, if 
\begin{equation}
    c_1=-\frac{3}{2 G^2 M^2}-2 \Lambda\,.
    \label{eq:c_1}
\end{equation}
Now setting $\gamma = 1$, $n=1$, and $\Lambda=0$, the obtained solution~\eqref{fR_reconstructed} corresponds to the standard Hilbert-Einstein Lagrangian resulting in the Schwarzschild spacetime.
\par Next, we search for an effective Lagrangian in the perturbation regime of the Bekeinstein-Hawking entropy that can imitate Eq.~\eqref{fR_reconstructed}. Now we expand up to the first-order Taylor series around $n=1$ to yield
\begin{align}\label{eq:multiv_Taylor}
\mathfrak{f(R)} \approx{} & c_1 + \left( \frac{3}{2G^2M^2} + \mathfrak{R} \right) \nonumber \\
& \times \left[ 1 + (n-1) \ln \left\{ \frac{2GM}{L_\gamma} \sqrt{1 + \frac{2}{3}G^2M^2\mathfrak{R}} \right\} \right]
\end{align}
This can be further reformed as
\begin{align} \label{compactfR}
\mathfrak{f(R)} \approx {} & c_1 + c_2 (1 + c_3 \mathfrak{R}) \nonumber \\ 
& \times \left[ 1 + (n-1) \ln\frac{2GM}{L_\gamma}+ \frac{(n-1)}{2} \ln(1 + c_3 \mathfrak{R})\right]\,,
\end{align}
where we denote
\begin{subequations}
    \begin{align}
\label{c2}
    c_2&\equiv \frac{1}{c_3}\left[1+(n-1)\ln \frac{2GM}{L_\gamma}\right], \\
    \label{c3}
    c_3&\equiv\frac{2}{3}G^2M^2\,.
    \end{align}
\end{subequations}
Consecutively, one can intuit Eq.~\eqref{compactfR} concurs with the first-order extension around $\epsilon=0$  of the function
\begin{equation}
  \mathfrak{f(R)}= {} c_1 +c_2\left(1 + c_3 \mathfrak{R}\right)^{1+\epsilon}\,,
   \label{fR_reconstructed_final}
\end{equation}
where 
\begin{equation}
\label{neweps}
  \epsilon\equiv\frac{n-1}{2}\,.  
\end{equation}
Hence, finally, we can infer that the GMHE can successfully be deduced from a special form of modified gravity Lagrangian provided in Eq.~\eqref{fR_reconstructed_final}. 
One can note that for $\epsilon=0$ (or equivalently $n=1$), $c_2=1/c_3$. Therefore, the standard $\mathfrak{f(R)}=\mathfrak{R}-2\Lambda$ gravitational model can be retrieved upon using of Eq.~\eqref{eq:c_1}.
\par In order to check consistency, one can plug Eq.~\eqref{fR_reconstructed_final} into Eq.~\eqref{entropy_f(R) horizon} by exploiting the definitions~\eqref{c2}, \eqref{c3} and \eqref{neweps}, and expanding to the first order around $n=1$ obtains
\begin{align}
S \approx {} & 4\pi GM^{2} \left( 1 + \frac{2}{3}G^{2}M^{2}\mathfrak{R}_{0} \right) \nonumber \\
& \times \left[1 + (n-1) \ln \left( \frac{2GM}{L_\gamma} \sqrt{1 + \frac{2}{3}G^{2}M^{2}\mathfrak{R}_{0}} \right) \right]\, .
\end{align}
This can be rewritten in a simple form as a function of horizon area:
\begin{equation}\label{eq:S_new}
   S \approx \frac{\mathcal{A}_{hor}}{4G} \left[ 1 + \frac{n-1}{2} \ln \left( \frac{\mathcal{A}_{hor}}{4\pi L^2_\gamma} \right) \right]\,.
\end{equation}
It is straightforward to establish that Eq.~\eqref{eq:S_new} conforms with the first-order Taylor series expansion around $n=1$ of the GMHE~\eqref{GS_entropy_area}. Consequently, this confirms the effectiveness of the Iyer-Wald method as well as gives validation to our obtained result \eqref{fR_reconstructed_final}.
\par It is noteworthy that akin logarithmic corrections to the horizon are prevalent in quantum entropic theories of gravity~\cite{banerjee2011logarithmic,kaul2000logarithmic,carlip2000logarithmic}. In this sense, this analysis could help for a comprehensive understanding of the essence of gravity and the spacetime fabric. In this context, it is also important to note that the leading-order corrections are proportional to $\ln(\mathcal{A}_{hor}/L_{Pl}^2)$, where $L_{Pl}=\sqrt{G}$ is the Planck length. This fundamentally relates $L_{\gamma}\equiv L_{Pl}$ in this theory.
%%%%%%%%%%%%%%%%%%%%%%%%%%%%%%%%%%%%%%%%%%%%%%%%%%%%%%%%%%%%%%%%%%%%%%%%%%%%%%%%%%%%%%%%%%%%%%%%%%%%%%%%%%%%%%%%%%%%%%%%%
\section{Physical consequences}
\label{PhysImpl}
\par The newly introduced modified gravity Lagrangian~\eqref{fR_reconstructed_final} encripts corrections to the standard Einstein-Hilbert action through a tiny parameter $\epsilon$. Surprisingly, this result may be viewed as an extension of GR by means of a Lagrangian  $\mathfrak{L}\propto \mathfrak{R}^{1+\epsilon}$, which is first presented in Refs.~\cite{clifton2005power,clifton2014erratum}, where authors have studied cosmological and weak-field properties of gravity. Moreover, several other applications have been subsequently investigated in relation to BH physics~\cite{de2023exploring,d2024geometric} and gravitational waves~\cite{capozziello2008higher}. In this section, we shall discuss the implications of our findings for cosmology and BH physics.
%%%%%%%%%%%%%%%%%%%%%%%%%%%%%%%%%%%%%%%%%%%%%%%%%%%%%%%%%%%%%%%%%%%%%%%%%%%%%%%%%%%%%%%%%%%%%%%%%%%%%%%%%%%%%%%%%%%%%%%%%%%%%%%%%%%%%%%%%%%%
\subsection{Cosmological implications}
\par Using the synthesis of the light elements, like lithium-$7$, helium-$4$, and deuterium, in a generalized cosmological framework, the value of the dimensionless parameter $\epsilon$ has been constrained in a study presented in Refs.~\cite{clifton2005power,clifton2014erratum}. They found the bound $-0.017<\epsilon<0.0012$. Using such a result and from the expression~\eqref{neweps}, we are able to impose upon $n$, the constraint
\begin{equation}
\label{CliftBar}
0.966\lesssim n \lesssim1.0024\,. 
\end{equation}
Remarkably, this bound is compatible with the $95\%$ confidence limit on the GMHE-inspired modified cosmology obtained from DESI DR$2$ BAO observational data~\cite{luciano2025modified}. The constraint on $n$ imposed from a study in Ref.~\cite{luciano2026baryogenesis} on GMHR-driven baryogenesis mechanism is also consistent with the aforementioned bound.
%%%%%%%%%%%%%%%%%%%%%%%%%%%%%%%%%%%%%%%%%%%%%%%%%%%%%%%%%%%%%%%%%%%%%%%%%%%%%%%%%%%%%%%%%%%%%%%%%%%%%%%%%%%%%%%%%%%%%%%%%
\subsection{Black hole thermodynamic stability}
\par In the realm of BH thermodynamics, particular emphasis is given to the notion of stability, which fundamentally depends on the thermodynamic variables of the horizon \cite{davies1978thermodynamics}. In this context, the study of local thermodynamic stability of BH is closely associated with the requirement that the heat capacity $\mathcal{C}$ be positive. Furthermore, any singularities or change in the sign of $\mathcal{C}$ indicate the onset of a phase transition and breakdown of thermodynamic equilibrium.
\par In the conventional theory of GR, the Schwarzschild solution results in BHs that are unstable in a local sense. To comprehend this, we recall that Hawking's temperature can be obtained from~\cite{czinner2016renyi}
\begin{equation}
    T_H=\left(\partial_M S\right)^{-1}\,,
\end{equation}
which yields for the heat capacity~\cite{czinner2016renyi}
\begin{equation}
\label{HC}
   \mathcal{C}=-\frac{(\partial_M S)^2}{\partial_M^2 S}\,.
\end{equation}
Exploiting the Bekenstein-Hawking area law $S_{BH}=\mathcal{A}_{hor}/(4G)$, one get $\mathcal{C}=-8\pi G M^2<0$. The negative sign of $\mathcal{C}$ signifies the fact that the Schwarzschild BH possesses local thermodynamic instability. This corresponds to the BH heating up due to radiating energy. However, it is worth mentioning that the Schwarzschild BH can show stability through cavity confinement~\cite{york1986black} or consideration of a large enough cosmological constant~\cite{hawking1983thermodynamics} while staying within the conventional framework of Boltzmann-Gibbs statistics.
\par Our next objective is to examine the stability condition in the GMHE scenario. For this occasion, we recast Eq.~\eqref{GS_entropy_area} as 
\begin{equation}\label{NewTs}
S_{GS} = \gamma \frac{2\pi}{G} \left( \frac{n}{n+1} \right) r_{hor}^{n+1}\,.
\end{equation}
In order to calculate the heat capacity for the GMHE~\eqref{NewTs}, we shall utilize the horizon radius given in Eq.~\eqref{rh}, which has been obtained in the perturbation regime surrounding the Schwarzschild BH solution. Accordingly, to maintain consistency, we should only consider the leading order in $\mathfrak{R}_0$, resulting in
\begin{align}\label{eq:C}
\mathcal{C} &= - 2^{n+2} \pi \gamma G^n M^{n+1} \left[ 1 + \frac{G^2 M^2 (n+3)(n-2) \mathfrak{R}_0}{3n} \right] + \mathcal{O}[\mathfrak{R}_0^2]\,.
\end{align}
One can notice that, for $n\rightarrow 1$, $\gamma\rightarrow 1$ and $\mathfrak{R}_0\rightarrow 0$, we retrieve the standard expression of heat capacity of Schwarzschild's BH, $\mathcal{C}= -8\pi G M^2$.
\par Therefore, we note that the GMHE significantly influences the BH thermodynamic structure. This could be either stable or unstable, based on the value of the entropic exponent parameter $n$. To determine the analytical range of the value of $n$ for which that heat capacity becomes positive, we expand Eq.~\eqref{eq:C} to the first order around $n=1$. By performing this expansion, it yields
\begin{align}\label{HeatCap}
\mathcal{C} \approx {} & -8\pi G M^2 \bigg[ \left(1 - \frac{4}{3} G^2 M^2 \mathfrak{R}_0\right) + (n-1) \bigg\{(1 + G^2 M^2 \mathfrak{R}_0) \nonumber \\
& + \left(1 - \frac{4}{3} G^2 M^2 \mathfrak{R}_0\right) \left( \ln\frac{2GM}{L_\gamma} - 1 \right) \bigg\}\bigg] + \mathcal{O}[(n-1)^2]\,. 
\end{align}
Accordingly, demanding $\mathcal{C}>0$ from Eq.~\eqref{HeatCap}, one obtains
\begin{equation}\label{n_condition}
n \lesssim  1 - \frac{1}{\ln \left( \frac{2GM}{L_\gamma} \right)} + \frac{7 G^2 M^2 \mathfrak{R}_0}{3 \left[ \ln \left( \frac{2GM}{L_\gamma} \right) \right]^2}
\end{equation}
under the condition\footnote{This preserves a valid perturbative regime ($n\approx 1$) and prevents inducing unphysical correction term.}
\begin{equation}
\label{M}
    M\gtrsim M_0\equiv\frac{L_\gamma}{2 G } \,,
\end{equation}
where the subdominant contributions in $\mathfrak{R}_0=4\Lambda\simeq 10^{-52}\,\text{m}^{-2}$ has been neglected. Undoubtedly, despite the value of mass of BH within the observational mass span, $M_\odot\lesssim M\lesssim10^{11} M_\odot$ ($M_\odot$ is the solar mass), the last component on the right side of the inequality \eqref{n_condition} can be ignored in comparison to the first two. It becomes apparent that the solution~\eqref{n_condition} indicates the only physically feasible limit. In reality, if considering the typical length scale $L_\gamma \sim L_{Pl} = \sqrt{G}$ as discussed in the end of previous Sec. \ref{sec:f(R)_BHs}, we get $M_0\sim M_{Pl}/2\simeq 10^{-8}\,\mathrm{kg}$, making the condition~\eqref{M} plausible for any observational BH within the specified mass range.
\par A few comments on the above analysis as follows. At first, we point out that the GMHE setup enables us to address the thermodynamic instability of Schwarzschild BHs by choosing a suitable condition on the entropic exponent parameter. Moreover, the upper limit on $n$ is influenced by changes in the BH mass, with lower $M$ values leading to a greater deviation from the 
holographic entropy scaling ($n = 1$). In particular, when considering the entire range of observed BH masses, it is found that 
\begin{equation}
\label{dunmezzo}
    n\lesssim  0.9886 \,.
\end{equation}
Interestingly, this constraint coincides with the range specified in Eq.~\eqref{CliftBar}, although evaluated in a different scenario.
%%%%%%%%%%%%%%%%%%%%%%%%%%%%%%%%%%%%%%%%%%%%%%%%%%%%%%%%%%%%%%%%%%%%%%%%%%%%%%%%%%%%%%%%%%%%%%%%%%%%%%%%%%%%%%%%%%%%%%%%%
\section{Summary \& Conclusions}\label{Concl}
\par In this letter, we investigated the connection between geometry and thermodynamics in the context of the GMHE, a two-parameter extension of standard Bekenstein-Hawking entropy. In so doing, we have considered BH solutions in the framework of $\mathfrak{f(R)}$ gravity, by making use of Iyer-Wald's Lagrangian formalism in the GMHE setup. 
\par We begin with the BH line element in $\mathfrak{f(R)}$ gravity, where we get the modifications to the Schwarzschild radius under small perturbations around GR. After that, by calculating the area of the BH event horizon and contrasting the resulting entropy with the expression of GMHE, we obtained the analytic expression of $\mathfrak{f(R)}$ in terms of model parameters ($\gamma$ and $n$) and an arbitrary constant. The constant was chosen by ensuring that our results perfectly aligned with the Schwarzschild-de Sitter solution in the standard entropic case $\gamma=n=1$.
\par We reconstructed an effective Lagrangian by reformulating $\mathfrak{f(R)}$ in terms of a dimensionless parameter $\epsilon\equiv\frac{n-1}{2}$, which quantifies slight deviations from GR. Specifically, we exhibit $\mathfrak{f(R)} = c_1+c_2\left(1 + c_3 \mathfrak{R}\right)^{1+\epsilon}$, which resembles to familiar model $\mathfrak{f(R)}\propto \mathfrak{R}^{1+\epsilon}$~\cite{clifton2005power,clifton2014erratum}. Further, considering the limitation on $\epsilon$ obtained from the synthesis of light elements~\cite{clifton2005power,clifton2014erratum}, we discussed the physical significances of our findings. In particular, we deduced the constraint on $n$: $0.966 \lesssim n\lesssim 1.0024$, which is in agreement with recent DESI DR$2$ BAO observations~\cite{luciano2025modified} and analysis of baryogenesis mechanism~\cite{luciano2026baryogenesis} on GMHE-inspired cosmology. In addition, we explored the problem of thermodynamic instability of Schwarzschild BHs, showcasing that the GMHE might offer a potential resolution under certain conditions.
\par Notably, logarithmic-like corrections to the horizon entropy arise frequently in quantum gravity theories~\cite{banerjee2011logarithmic,kaul2000logarithmic,carlip2000logarithmic}, with leading-order terms proportional to $\ln(\mathcal{A}_{hor}/L_{Pl}^2)$. This aligns our framework with quantum entropic theories by fundamentally identifying the length scale $L_{\gamma}$ with the Planck length $L_{Pl}$. This could help for a comprehensive understanding of the nature of gravity and the spacetime fabric. It is worth mentioning that the impact of massless matter and quantum gravity on the thermodynamic configurations of Schwarzschild BHs has also been explored in Ref.~\cite{el2017quantum}, indicating that quantum gravity effects could solve the instability issue. It would be discerning to associate such an outcome with our result and possibly revisit the root cause of the GMHE-inspired modified gravity model in the quantum theoretical framework. The results concerning our article suggest a potential correction to GR on BH physics as well as on cosmological scales and within the context of BH physics to account for the observed phenomenology.
\par In conclusion, it is important to highlight the fact that the present study may have consequences that extend beyond the specified context we discussed. In fact, GMHE can be represented by Eq.~\eqref{eq:S_new} in the operational regime $|n-1|\ll 2$. Alike logarithmic corrections to the horizon scaling frequently emerge in quantum gravity theories, including loop quantum gravity, string theory, the generalized uncertainty principle, and AdS/CFT correspondence. Given the strong connections among quantum mechanics, statistical mechanics, gravitation, and cosmology, our study approaches in contributing to a profound understanding of gravity and the nature of spacetime from both thermodynamics and information theory viewpoints. Research on this topic is already in progress and will be discussed somewhere else.
%%%%%%%%%%%%%%%%%%%%%%%%%%%%%%%%%%%%%%%%%%%%%%%%%%%%%%%%%%%%%%%%%%%%%%%%%%%%%%%%%%%%%%%%%%%%%%%%%%%%%%%%%%%%%%%%%%%%%%%%%%%%%%%%%%%%
%%%%%%%%%%%%%%%%%%%%%%%%%%%%%%%%%%%%%%%%%%%%%%%%%%%%%%%%%%%%%%%%%%%%%%%%%%%%%%%%%%%%%%%%%%%%%%%%%%%%%%%%%%%%%%%%%
\section*{Data availability statement}
No data was used for the research described in the article.
%%%%%%%%%%%%%%%%%%%%%%%%%%%%%%%%%%%%%%%%%%%%%%%%%%%%%%%%%%%%%%%%%%%%%%%%%%%%%%%%%%%%%%%%%%%%
\section*{Declaration of competing interest}
The authors declare that they have no known competing financial interests or personal relationships that could have appeared to influence the work reported in this paper.
%%%%%%%%%%%%%%%%%%%%%%%%%%%%%%%%%%%%%%%%%%%%%%%%%%%%%%%%%%%%%%%%%%%%%%%%%%%%%%%%%%%%%%%%%%%%%%%%%%%%%%%%%%%%%%%%%%%%%%%%%%%%%%%%%%%%
\begin{acknowledgments}
SM acknowledges the Government of West Bengal, India, for providing the State-funded Senior Research Fellowship (SRF).
\end{acknowledgments}
%%%%%%%%%%%%%%%%%%%%%%%%%%%%%%%%%%%%%%%%%%%%%%%%%%%%%%%%%%%%%%%%%%%%%%%%%%%%%%%%%%%%%%%%%%%%%%%%%%%%%%%%%%%%%%%

\newpage
\bibliography{reference}

\end{document}